\documentclass[12pt,letterpaper]{article}
\usepackage[utf8]{inputenc}
\usepackage{tabularx}
\usepackage{array}
\usepackage{placeins}
\usepackage{longtable}
\usepackage{newtxtext,newtxmath}
\usepackage{lipsum}
\usepackage{amsmath}
\usepackage{amsfonts}
\usepackage{graphicx}
\usepackage{cite}
\usepackage{ulem}
\usepackage{caption}
\usepackage[top=1in,bottom=1in,left=1in,right=1in]{geometry} 
\usepackage{float}
\usepackage{hyperref}
\usepackage{booktabs}
\usepackage[export]{adjustbox}
\usepackage{float}
\usepackage{multirow}
\usepackage{pifont}
\usepackage[ ]{authblk}
\usepackage{chemarrow}
\usepackage{comment}
\usepackage{makecell}
\UseRawInputEncoding
\hypersetup{
    colorlinks=true,
    linkcolor=blue,
    filecolor=magenta,      
    urlcolor=blue,
    pdftitle={Overleaf Example},
    pdfpagemode=FullScreen,
}
\usepackage{multicol}
\begin{document}

%
%
\title{Physics-Informed Electrochemical Model of Cathodic Corrosion in Alkaline Media}
%
%

%
%
\author[1]{Auronno Ovid Hussain}
\author[1]{Abdul Ahad Mamun}
\author[1]{Faysal Rahman}
\author[*,1]{Muhammad Anisuzzaman Talukder}
\affil[1]{\small{Department of Electrical and Electronic Engineering \\

Bangladesh University of Engineering and Technology, Dhaka 1205, Bangladesh}} 
\affil[*]{\small{\it{anis@eee.buet.ac.bd}}}
%
%

\date{ }
\maketitle
\sloppy

%
%
\begin{abstract}
Electrochemical corrosion significantly reduces the durability of electrodes in water electrolyzers, adversely affecting hydrogen (H$_2$) production and cell efficiency. Current theoretical models inadequately assess corrosion behaviors in alkaline water electrolyzers. To address this, we developed a physics-informed electrochemical corrosion model evaluating the corrosion characteristics of cathodes in alkaline systems, accounting for factors such as exchange current density ($J_0$), redox potential ($E_0$), Gibbs free energy of hydrogen adsorption ($\Delta G_{\rm H}$), electrolyte concentration ($C$), system pressure ($P$), and temperature ($T$). The model calculates metrics including corrosion potential ($E_{\rm corr}$), corrosion current density ($J_{\rm corr}$), and corrosion rate ($C_R$). Our findings from potentiodynamic polarization indicate that gold (Au) shows the highest durability, while copper (Cu) and nickel (Ni) are promising cost-effective alternatives. This work enhances the understanding of corrosion dynamics, contributing to the design of more efficient electrolyzer cells for hydrogen production.
\end {abstract}
%
%

%
%

%
%
\section{Introduction}

Global energy demand is increasing due to a growing world population, rapid industrialization, and high power consumption driven by advancing technology \cite{sun2025lignin}. Addressing energy scarcity is a significant challenge, while mitigating global warming is one of the major issues of the $21^{\rm st}$ century \cite{ren2024recent}. Hydrogen (H$_2$) gas is a promising alternative to fossil fuels due to its high energy content and zero carbon emissions \cite{zhang2023electrochemical}. Although various chemical processes and technologies can be employed to produce H$_2$ fuel, water electrolysis offers a favorable technique due to its decarbonized way of generating H$_2$ \cite{hassan2024recent}. This approach also offers several advantages, including the production of high-purity hydrogen, its non-toxic nature, and the ability to use a wide range of power sources \cite{el2023hydrogen}. The main components of water electrolyzer cells include electrodes, catalysts, ion exchange membranes, electrolytes, and stack modules \cite{lee2021multidimensional}. Among these, the electrode is the most critical part of the electrolyzer cell, as it facilitates charge transfer to the electrolyte, producing H$_2$ and oxygen (O$_2$) gases through the hydrogen evolution reaction (HER) and the oxygen evolution reaction (OER), respectively \cite{liu2022rational}.

The catalytic activity and chemical stability of electrodes are crucial for effective and sustainable H$_2$ production during water electrolysis \cite{hou2022strategies}. However, various factors, such as corrosion, mechanical instability, and electrolyte contamination, can significantly diminish electrode performance \cite{arun2025challenges, sun2023electrochemical}. Corrosion is particularly critical as it leads to the degradation of metal electrodes in a water electrolyzer cell \cite{li2022corrosion}. Electrodes can corrode through both chemical and electrochemical processes \cite{perez2024electrochemical}. Chemical corrosion occurs through interactions between the electrode's surface and the surrounding environment without charge carrier transfer. On the other hand, electrochemical corrosion is a more complex process that involves the transfer of charge carriers in electrolytic conditions. According to the electrochemical theory, electrochemical corrosion occurs when two dissimilar metals come into contact with a conductive aqueous solution \cite{kumar2024advancements}. Electrochemical corrosion is one of the most common forms of corrosion processes in water electrolyzer cells due to the continuous charge transfer at the interface between the electrode and the electrolyte \cite{wallnofer2024review}. 

Recent advancements in corrosion modeling, such as mixed potential theory (MPT), finite element methods, and machine learning, have been developed to investigate the electrochemical corrosion of electrode materials in various electrolytes \cite{li2025integration}. Wagner et al.~developed MPT to assess the degradation of metals in electrolyte environments \cite{wagner2006interpretation}. By incorporating constant exchange current density ($J_0$), Tafel slope for HER, and activity of metal ions into the MPT, Lazzari et al.~derived the Tafel-Piontelli model, which can predict the corrosion behavior of active metals in strongly acidic medium \cite{messinese2022tafel}. Subsequently, Messinese et al.~modified the model by including the dependency on temperature ($T$) and electrolyte pH \cite{messinese2022tafel}. However, both of these models focused solely on the corrosion caused by the HER. They did not consider the corrosion associated with metal oxide formation under alkaline conditions, assuming that the anodic Tafel slope of metals is approximately zero. Additionally, these models do not measure the corrosion of non-active metals, including copper (Cu), silver (Ag), gold (Au), platinum (Pt), and iridium (Ir) in both acidic and alkaline conditions \cite{messinese2022tafel}. Later, the Tafel-Piontelli model was adjusted to incorporate non-zero anodic Tafel slopes and $J_0$ for metal dissolution, allowing it to predict the corrosion behavior of active metals in weak acids \cite{messinese2024predictive}. Nevertheless, the applicability of this improved model remains limited to carbon-based materials, such as carbon steel, which impedes its utilization for noble and non-noble metal electrodes. This limitation of the adjusted Tafel-Piontelli model suggests that further model improvement is required to fully understand the mechanisms of electrochemical corrosion across a broader range of materials and electrolyte conditions. 

Finite element-based corrosion models have been used to examine the electrochemical corrosion of galvanic couples used in industrial applications \cite{snihirova2019galvanic, dong2025study}. However, due to the high computational cost, inherent modeling complexities, and reliance on numerous empirical parameters, these models cannot reliably predict corrosion rates across different electrochemical systems \cite{snihirova2019galvanic}. Recently, machine learning and artificial intelligence-driven corrosion models have been explored to analyze the electrochemical corrosion of metals \cite{zhao2024efficient, madhavan2025data}. Despite their innovations, these data-driven statistical models are limited by the insufficient incorporation of electrochemical theories of corrosion, which hinders their predictive accuracy and widespread application. Therefore, developing a robust electrochemical corrosion model is essential for assessing the stability and durability of metal electrodes in water electrolyzer cells. The commercially available alkaline electrolyzer cells play a crucial role in producing H$_2$, suggesting that it is necessary to investigate electrode corrosion in alkaline medium \cite{wang2024bridging}. Our investigation can predict optimal operating lifespans and long-term stability, providing essential design insights to select appropriate electrode materials for alkaline water electrolyzer cells under varying electrochemical conditions.

In this work, we have developed an advanced electrochemical corrosion model to investigate the steady-state corrosion of metal electrodes used in alkaline water electrolyzers. This model advances the existing approaches by incorporating the effects of multiple oxidation states of metals, mass transfer limitations, and H$_2$ adsorption, which are crucial for understanding the electrochemical corrosion mechanism of metal electrodes. Additionally, empirical parameters for electrolyte concentration ($C$) and system pressure ($P$) have been incorporated in the model to enhance its applicability across various electrochemical environments.

Electrochemical analyses, such as cyclic voltammetry (CV) and potentiodynamic polarization techniques, have been performed to validate the model. Qualitative insights into the electrochemical corrosion process have been described by examining the changes in double-layer capacitance and oxidation peak currents at varying electrolyte concentrations. The empirical parameters of our model are measured using the potentiodynamic polarization data of Cu plates. The model has been used to calculate key corrosion parameters, including corrosion potential ($E_{\rm corr}$), corrosion current density ($J_{\rm corr}$), and corrosion rate ($C_R$) of both noble and non-noble metal electrodes, and these results are benchmarked against experimental data. The statistical metrics reveal a strong correlation between model predictions and experimental results, indicating the model's accuracy. The model provides valuable information regarding the stability and durability of both noble and non-noble metal electrodes by comparing their electrodissolution. Therefore, the developed model can be a powerful tool for designing robust and sustainable alkali-based water electrolyzers for efficient H$_2$ production. 
%
%

\section{Electrochemical Corrosion Model for Water Electrolysis}
\subsection{Fundamental Principles}

Water electrolysis is an electrochemical process in which OER and HER occur at the anode and the cathode electrodes, respectively. However, metal dissolution is an undesirable reaction that occurs at the electrode's surface. To develop the electrochemical corrosion model for metal electrodes in alkaline water electrolysis, it is necessary to consider both HER and metal (M) dissolution. The corresponding redox reactions are as follows \cite{mamunadvancing}
\begin{subequations}
\begin{equation}
    2n\mathrm{H_2O} + 2n \mathrm{e^{-}} \rightarrow n\mathrm{H_2} + 2n\mathrm{OH^{-}},
    \label{eq:1(a)}
\end{equation}
\begin{equation}
    2\mathrm{M} \rightarrow 2\mathrm{M}^{n+} + 2n \mathrm{e^{-}},
    \label{eq:1(b)}
\end{equation}
\end{subequations}
where $n$ is the oxidation number of the metal electrode. 

\begin{figure}[ht]
    \centering
    \includegraphics[width =\linewidth]{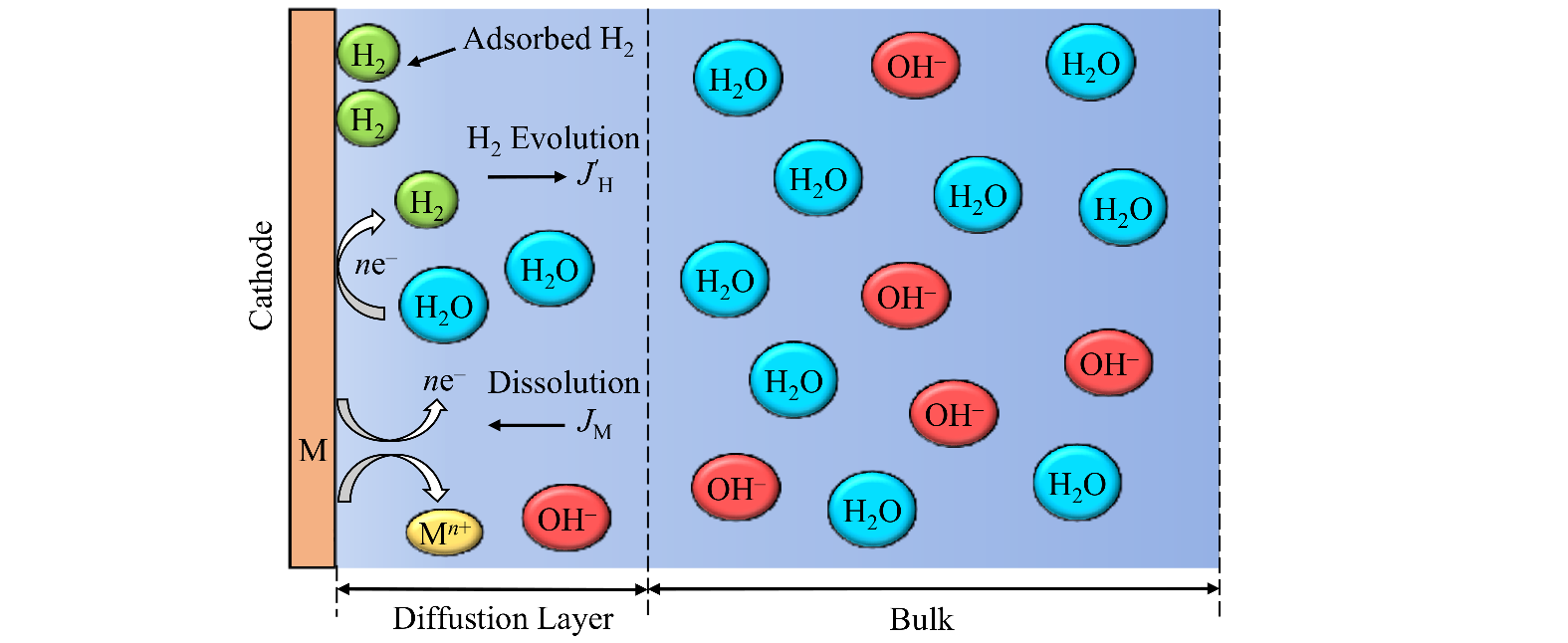}
    \caption{Schematic illustration of the electrochemical corrosion of metal (M) cathodes in an alkaline electrolyte. $J_{\rm M}$ and $J'_{\rm H}$ represent the currents due to metal dissolution and H$_2$ evolution reactions, respectively.}
    \label{Figure 01}
\end{figure}

In an alkaline medium, H$_2$ is produced from a water molecule (H$_2$O) by a reducing reaction. This reduction process causes electron transfer from the metal surface to the electrolyte, resulting in metal dissolution. These reactions, occurring on the cathode surface, are shown in Fig.~\ref{Figure 01}. 
Various redox reactions with distinct potentials occur at the cathode surface, leading the system to stabilize at a new redox potential where the reaction kinetics are balanced \cite{wagner2006interpretation}. This stabilized potential is referred to as $E_{\rm corr}$, which qualitatively describes the electrochemical corrosion mechanisms and indicates the tendency of electrons to detach from the metal surface. When $E_{\rm corr}$ decreases, the electron transfer rate toward the electrolyte increases, resulting in more electrochemical corrosion \cite{tait2018electrochemical}. The metal dissolution current, $J_{\rm corr}$, at $E_{\rm corr}$ provides a quantitative assessment of electrochemical corrosion and describes the involved kinetic processes \cite{tait2018electrochemical}. A higher value of $J_{\rm corr}$ is associated with an increased rate of metal degradation. 

\subsection{Model Development and Assumptions}

The theoretical model for electrochemical corrosion is based on three main assumptions: (i) the electrochemical reactions occurring on the metal surface can be divided into multiple oxidation and reduction processes; (ii) charge conservation is maintained at the electrode surface; and (iii) the effects of O$_2$ crossover losses and the O$_2$ reduction reaction (ORR) are not considered at the cathode surface, as standard water electrolyzers use membranes that prevent O$_2$ diffusion toward the cathode \cite{wagner2006interpretation, yang2022anion}. The relationship between the current density ($J$) and the applied potential ($E$) of a water electrolyzer cell is governed by the Butler-Volmer equation expressed as \cite{dickinson2020butler}
\begin{equation}
    J = J_0 \left[\exp\left(\frac{E-E_0}{b_a}\right) - \exp\left(-\frac{E-E_0}{b_c}\right)\right],
    \label{eq:2}
\end{equation}
where $E_0$ is the standard redox potential of water splitting and $b_a$ and $b_c$ are the anodic and cathodic Tafel slopes, respectively. The Tafel slopes, $b_a$ and $b_c$, can be estimated using \cite{dickinson2020butler}
\begin{subequations}
    \begin{equation}
        b_a = \frac{RT}{\alpha z F},
        \label{eq:3(a)}
    \end{equation}
    \begin{equation}
        b_c = \frac{RT}{\left( 1 - \alpha \right) z F},
        \label{eq:3(b)}
    \end{equation}
\end{subequations}
where $R$, $\alpha$, $z$, and $F$ represent the molar gas constant, the reaction transfer coefficient, the number of electrons transferred, and the Faraday constant, respectively. The reaction transfer coefficients are usually assumed as $0.5$ \cite{dickinson2020butler}.

The metal dissolution primarily occurs as an anodic process, since it is kinetically unfavorable in reducing electrochemical environments \cite{li2021oxidative}. Thus, the cathodic term of Eq.~\eqref{eq:2} can be neglected, and the metal dissolution current density ($J_{\rm M}$) can be written as
\begin{equation}
    J_{\rm M} = J_{\rm 0,M} \left[\exp\left(\frac{E-E_{\rm 0,M}}{b_{\rm M}}\right)\right],
    \label{eq:4}
\end{equation}
where the subscript $\rm M$ stands for the metal dissolution reaction. Unlike metal oxidation, HER is fundamentally a cathodic process, and the HER current density, $J_{\rm H}$, can be derived from Eq.~\eqref{eq:2} given by
\begin{equation}
    J_{\rm H} = -J_{\rm 0,H} \left[\exp\left(-\frac{E - E_{\rm 0,H}}{b_{\rm H}}\right)\right],
    \label{eq:5}
\end{equation}
where the subscript $\rm H$ represents the HER. 

The concentration of metal ions in the electrolyte is negligible, and the system pressure of the water electrolyzer cell does not influence the properties of the metal \cite{perez2024electrochemical, scholz2017wilhelm}. As a result, $E_{\rm 0,M}$ remains unaffected by changes in system pressure and electrolyte concentration. In contrast, $E_{\rm 0,H}$ is impacted by both electrolyte concentration and system pressure \cite{horvath2013ph}. In an alkaline medium, $E_{\rm 0,H}$ can be estimated using the Nernst equation \cite{scholz2017wilhelm}
\begin{equation}
    E_{\rm 0,H} = -\frac{RT}{z_{H_2}F} \log\left( PC_{\rm OH^-}^2 \right),
    \label{eq:6}
\end{equation}
where $z_{\rm {H_2}}$, $C_{\rm OH^-}$, and $P$ are the number of electrons transferred during the HER process, concentration of $\rm OH^-$ ions in the electrolyte, and the system pressure, respectively.

In addition to $J_{\rm {M}}$ and $J_{\rm {H}}$, a constant leakage current ($J_{\rm {leak}}$) flows through the interface between the metal and the electrolyte \cite{shen2011concise}. $J_{\rm {leak}}$ results from the charging and discharging effects in the capacitive double layer that forms at the electrode-electrolyte interface. Both electrolyte concentration and system pressure influence the kinetics of metal dissolution \cite{huang2015selective, chen2011carbon}. An increase in electrolyte concentration reduces the resistance of the electrolyte, facilitating faster electron transfer toward the electrolyte, which ultimately increases metal dissolution \cite{huang2015selective}. Conversely, an increase in system pressure can disrupt the passivation layer on the metal surface, leading to the rapid oxidation of the metal electrodes \cite{chen2011carbon}. Moreover, metals can possess multiple oxidation states, each with its distinct redox potential and Tafel slope \cite{he2025review}. Therefore, we have modified Eq.~\eqref{eq:4} by introducing multiple exponential terms, which account for the different oxidation states of metals. The expression of $J_{\rm M}$ for multivalent metals is given by 
\begin{equation}
    J_{\rm M} = J_{\rm leak} + \sum_{k=1}^{L} J_{\rm 0,M} \left( \frac{Ce}{N_{\rm ref}} \right)^{\beta_{{\rm M},k}} \left( \frac{P}{P_{\rm ref}} \right)^{\gamma_{{\rm M},k}} \left[ \exp\left( \frac{E - E_{{\rm 0M},k}}{b_{{\rm M},k}} \right) \right],
    \label{eq:7}
\end{equation}
where $e$ is the number of replaceable OH$^-$ groups per molecule of the alkaline electrolyte, $N_{\rm ref}$ and $P_{\rm ref}$ are the reference normality and system pressure, respectively, $b_{{\rm M},k}$ and $E_{{\rm 0M},k}$ are the Tafel slope and metal redox potential for the $k^{\rm th}$ oxidation state, and $L$ is the total number of oxidation states. We have introduced $\beta_{{\rm M},k}$ and $\gamma_{{\rm M},k}$, which represent the empirical constants for concentration and pressure dependencies for the $k^{\rm th}$ oxidation state in metal dissolution, respectively. In this model, $N_{\rm ref}$ and $P_{\rm ref}$ are considered as $1$ N and $1$ atm, respectively.

Similar to metal dissolution, HER depends on both system pressure and electrolyte concentration \cite{tian2025progress, wei2023investigation}. According to Henry's law, increasing the system pressure enhances the dissolution of H$_2$. However, higher electrolyte concentration improves conductivity, making the reaction kinetics for hydrogen evolution more favorable. H$_2$ produced during the reaction adsorbs onto the metal surface, covering catalytic sites and subsequently leading to sluggish kinetics of HER \cite{jang2021numerical}. The Langmuir adsorption model provides a surface coverage factor ($\theta$), which is represented by \cite{mamun2024enhancing}
\begin{equation}
    \theta = \frac{\exp\left( \frac{-\Delta G_{\rm H}}{k_{\rm B} T} \right)}{1 + \exp\left( \frac{-\Delta G_{\rm H}}{k_{\rm B} T} \right)},
    \label{eq:8}
\end{equation}
where $\Delta G_{\rm H}$ is the Gibbs free energy for H$_2$ adsorption and $k_{\rm B}$ is the Boltzmann constant. 

During the HER, the concentration of ions near the electrode surface is lower than the bulk concentration. As a result, ions must diffuse from the bulk to the surface, which causes mass transfer limitations and reduces the overall reaction kinetics for hydrogen evolution. The maximum current density observed in the diffusion-controlled region is known as the limiting current density, $J_{\rm lim}$, which can be expressed as \cite{chehayeb2019electrical}
\begin{equation}
    J_{\rm lim} = \frac{z_{{\rm H}_2}FCD_{{\rm H}_2}}{\delta},
    \label{eq:9}
\end{equation}
where $D_{\rm {H_2}}$ is the diffusion coefficient of $\rm {H_2}$ at the operating temperature and pressure, and $C$ and $\delta$ are the concentration of electrolyte and the diffusion layer thickness, respectively. Diard et al.~derived an expression for $\delta$ in low potential, which is given by \cite{diard2005diffusion} 

\begin{equation}
    \delta = 3.5 \sqrt{\frac{D_{{\rm H}_2}RT}{z_{{\rm H}_2}Fv} },
    \label{eq:10}
\end{equation}
where $v$ represents the scan rate of the electrochemical system. The adsorption and mass transfer models have been used to incorporate the effects of $J_{\rm leak}$, electrolyte concentration, system pressure, mass transfer limitations, and surface coverage in Eq.~\eqref{eq:5} \cite{wei2023investigation, tian2025progress, popov2010effect}. The term $J_{\rm H}$ is modified to $J'_{\rm H}$ as 
\begin{equation}
     J'_{\rm H} = - J_{\rm leak} - \frac{J_{\rm 0,H} (1 - \theta) \left( \frac{Ce}{N_{\rm ref}} \right)^{\beta_{\rm H}} \left( \frac{P}{P_{\rm ref}} \right)^{-\gamma_{\rm H}} \left[ \exp\left( -\frac{E - E_{0{\rm H}}}{b_{\rm H}} \right) \right]}{ 1 + \frac{J_{0,{\rm H}}}{J_{\rm lim}} (1 - \theta) \left( \frac{Ce}{N_{\rm ref}} \right)^{\beta_{\rm H}} \left( \frac{P}{P_{\rm ref}} \right)^{-\gamma_{\rm H}} \left[ \exp\left( -\frac{E - E_{0{\rm H}}}{b_{\rm H}} \right) \right]},
     \label{eq:11}
\end{equation}
where $\beta_{\rm H}$ and $\gamma_{\rm H}$ are defined as the empirical constants for concentration and pressure dependencies of HER. 

The total cathodic current can be calculated by summing the contributions from $J_{\rm M}$ and $J'_{\rm H}$. According to the charge conservation law, we can consider the cathodic current flow to be negligible at the corrosion potential, $E_{\rm corr}$. This phenomenon leads to
\begin{equation}
    J_{\rm M} + J'_{\rm H} \approx 0.
    \label{eq:12}
\end{equation} 
Eq.~\eqref{eq:12} is the governing equation for the $E_{\rm corr}$ for pure metals in alkaline medium. By utilizing the value of $E_{\rm corr}$ calculated from Eq.~\eqref{eq:12}, $J_{\rm corr}$ is measured by the following equation as given by
\begin{equation}
    J_{\rm corr} = J_{\rm leak} + \sum_{k=1}^{L} J_{0,{\rm M}} \left( \frac{Ce}{N_{\rm ref}} \right)^{\beta_{{\rm M},k}} \left( \frac{P}{P_{\rm ref}} \right)^{\gamma_{{\rm M},k}} \left[ \exp\left( \frac{E_{\rm corr} - E_{0{\rm M},k}}{b_{{\rm M},k}} \right) \right].
    \label{eq:13}
\end{equation}

Although Eqs.~\eqref{eq:12} and \eqref{eq:13} are correct for high concentrations, it is necessary to make adjustments to the governing equations at low concentrations. At low $C$, the ionic strength of the electrolyte is very low, leading to a large Debye length \cite{riad2020analysis}. As a result, the diffusion layer becomes thicker, and the double-layer capacitance formed at the electrode-electrolyte interface decreases. Therefore, $J_{\rm leak}$ can approach approximately zero. The OH$^{-}$ ions of the alkaline electrolyte are Lewis bases that function as weak stabilizing ligands \cite{ishii2009universal}. They provide transient stability to unstable metal ions via oxide formation \cite{starosvetsky2013features}. However, as the availability of ligands diminishes at lower electrolyte concentrations, the oxide formation becomes kinetically less favorable. Therefore, the unstable metal ions spontaneously oxidize or reduce to attain the most thermodynamically stable ionic configurations. This phenomenon allows us to eliminate the multi-oxidation states from Eq.~\eqref{eq:7}. The modified $J_{\rm corr}$ for low concentration can be written as
\begin{equation}
     J_{\rm corr} = J_{0,{\rm M}} \left( \frac{Ce}{N_{\rm ref}} \right)^{\beta_{{\rm M},1}} \left( \frac{P}{P_{\rm ref}} \right)^{\gamma_{{\rm M},1}} \left[ \exp\left( \frac{E_{\rm corr} - E_{0{\rm M},1}}{b_{{\rm M},1}} \right) \right].
    \label{eq:14}
\end{equation}

Eqs.~\eqref{eq:12} and \eqref{eq:13} represent the governing equations for analyzing the electrochemical corrosion of a metal electrode during alkaline water electrolysis. The mass lost in the corrosion process can be converted to $J_{\rm corr}$ and subsequently to $C_R$ using Faraday's law of electrolysis \cite{tait2018electrochemical}. The expression for $C_R$ can be given by
\begin{equation}
    C_R = \frac{MJ_{\rm corr}}{z_{\rm M}dF},
    \label{eq:15}
\end{equation}
where $M$, $z_{\rm M}$, and $d$ are the atomic mass, the valency, and the density of the electrode material, respectively. If a metal has multiple valencies, the most stable one is used in Eq.~\eqref{eq:15}.   

\subsection{Model Characteristics}

According to the governing Eqs.~\eqref{eq:12} and \eqref{eq:13}, the electrochemical corrosion of a metal cathode depends on its material properties and the system characteristics. The material properties include $J_0$, $E_0$, and $\Delta G_{\rm H}$. Moreover, $C$, $P$, and $T$ also affect the electrochemical corrosion of metals. The governing equations were simplified to analyze the optimal ranges of the empirical constants. We have fitted the advanced electrochemical corrosion model using the corrosion data of Cu. Therefore, the number of oxidation states has been considered as $2$, providing only two exponential terms for $J_{\rm M}$. The two reaction pathways of the metal dissolution process are governed by the same electrolyte environment and system properties, which influence the reaction kinetics. Therefore, we have considered that both $\beta_{\rm M,1}$ and $\beta_{\rm M,2}$ equal to $\beta_{\rm M}$. Similarly, it was assumed that $\gamma_{\rm M,1}$ and $\gamma_{\rm M,2}$, were equal to $\gamma_{\rm M}$. Monohydroxy electrolytes are traditionally used in alkaline water electrolyzer cells, allowing us to consider $e=1$ in the governing equations \cite{el2023hydrogen}. 

\begin{figure}
    \centering
    \includegraphics[width =\linewidth]{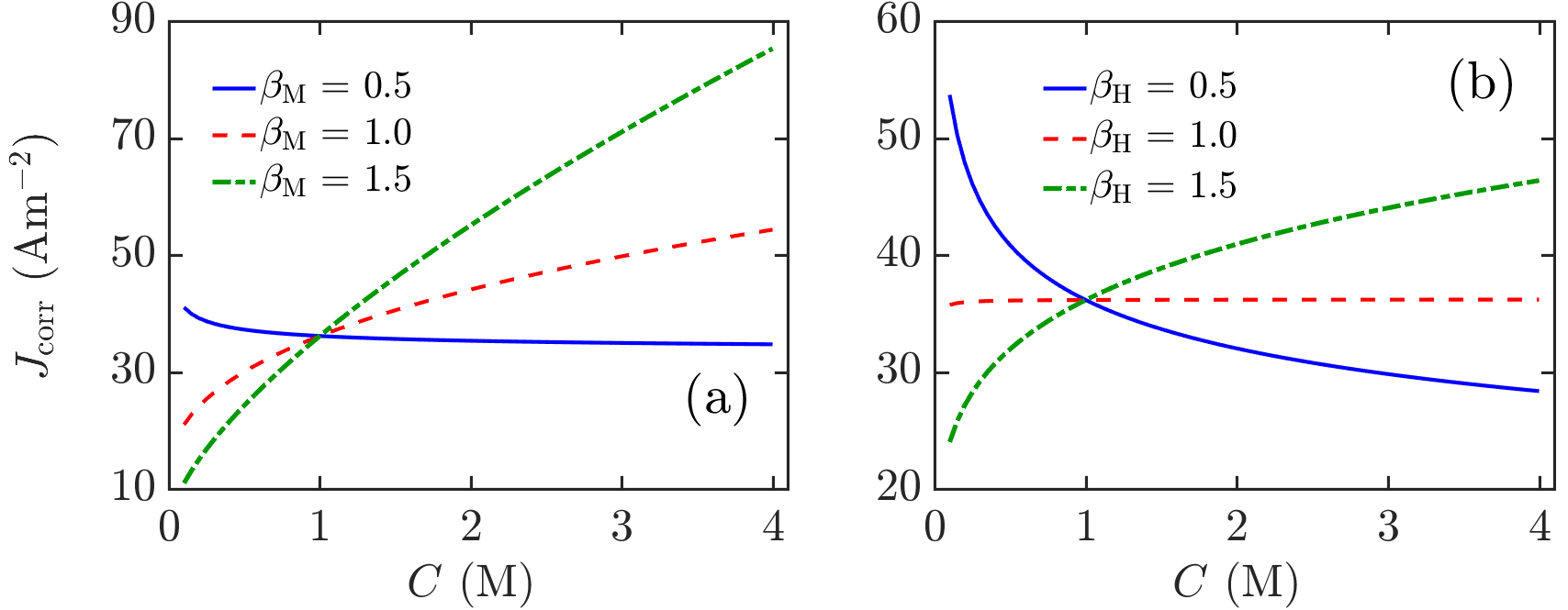}
    \caption{Impacts of (a) $\beta_{\rm M}$ and (b) $\beta_{\rm H}$ variation on corrosion current density ($J_{\rm corr}$). The parameters $\beta_{\rm M}$ and $\beta_{\rm H}$ are the empirical constants of concentration dependencies of metal dissolution and hydrogen evolution at metal cathodes in alkaline water splitting, respectively. A standard pressure ($P$) of 1 atm was considered.}
    \label{Figure 02}
\end{figure}

An increase in $C$ causes an increase in the ionic strength and conductivity, enhancing the overall reaction kinetics of metal dissolution. According to electrochemical theories and experimental trends, $J_{\rm corr}$ has a positive non-linear correlation with $C$ \cite{li2019electrochemical}. Therefore, we varied the empirical constants, $\beta_{\rm M}$ and $\beta_{\rm H}$, to find their optimal range for physically realizable solutions. To eliminate pressure dependencies, $P$ was considered $1$ atm \cite{li2020metallic}. Figure \ref{Figure 02}(a) suggested that $J_{\rm corr}$ showed different trends, according to a change in $\beta_{\rm M}$. When $\beta_{\rm M}\gtrsim 0.5$, $J_{\rm corr}$ increased non-linearly with $C$. This trend was consistent with the established theories and trends observed in experiments \cite{li2019electrochemical}. Thus, the value of $\beta_{{\rm M}}$ must be greater than $0.5$ for physically meaningful solutions. According to the results shown in Fig.~\ref{Figure 02}(b), $\beta_{\rm H}$ must be greater than $1$ to ensure valid solutions. 

As $P$ increases, the passivation layer formed on the cathode surface becomes damaged, leading to more spontaneous electron transfer toward the electrolyte, ultimately increasing the kinetics of electrochemical corrosion. To investigate the optimal ranges of both $\gamma_{\rm M}$ and $\gamma_{\rm H}$, we have varied them from $0.5$ to $1.5$. The $C$ was considered as $1$ M to exclude any concentration dependencies. Figure S1(a) illustrates how  $\gamma_{\rm M}$ affects the change of $J_{\rm corr}$ with $P$. When $\gamma_{{\rm M}}\gtrsim0.5$, $J_{\rm corr}$ showed an increasing trend with $P$. This increasing trend aligned with previously reported experimental results \cite{zhu2019effect}. Consequently, the constant $\gamma_{{\rm M}}$ must be $\gtrsim0.5$ for physically meaningful solutions. Similarly, analysis of Fig.~S1(b) indicates that $\gamma_{\rm H}$ must be less than $-0.5$ for the model to yield valid solutions. This increasing trend aligned with the fundamentals of electrochemistry and experimental results \cite{zhu2019effect}. Therefore, the constant $\gamma_{{\rm M}}$ has a lower bound of $0.5$. Meanwhile, $\gamma_{\rm H}$ must be less than $-0.5$, according to Fig.~S1(b), to obtain physically realizable solutions.

\section{Experimental Procedure}
\subsection{Material Preparation}

Industry-grade Cu sheets with a purity of 99.90\% were cut into $10 \times 10 \times 1$ mm-sized plates. The plates were cleaned and rinsed with deionized (DI) water, acetone, and ethanol. To prepare the electrode surfaces for electrochemical analysis, the plates were carefully polished with emery paper before each test. The DI water, sourced from the German Lab, had a conductivity of $<5.0$ {\textmu}Scm$^{-1}$, and a pH of 8. Acetone and ethanol were obtained from Sigma-Aldrich, while anhydrous potassium hydroxide (KOH) pellets were obtained from Merck for the preparation of alkaline solutions. All reagents used in this work had a purity of $\geq99.5$\%. 
\subsection{Electrochemical Measurements}

The experiment was carried out in $250$ mL sealed H-cells, using a standard three-electrode method. The Cu plate was utilized as the working electrode (WE). Platinum (Pt) plate and saturated Ag/AgCl were used as the counter electrode (CE) and the reference electrode (RE), respectively. The concentration of the alkaline electrolyte was varied from $0.25$ to $2.00$ M. The temperature was kept constant at $25$ $^{\rm o}$C. The room pressure was $1$ atm throughout the experiment.  

Corrtest CS350M electrochemical workstation was used to perform the electrochemical analyses, such as CV and potentiodynamic linear polarization resistance (LPR) measurements. The electrochemical measurements were conducted after the open-circuit potential (OCP) had stabilized. Both CV and LPR analyses were carried out from $-1.2$ to $+0.6$ V with a scan rate of $10$ mVs$^{-1}$. All voltages were referenced to the RE. The double-layer capacitance, $C_{\rm dl}$, was calculated in CS Studio Analysis using the CV curves. The potentiodynamic polarization technique was used for corrosion measurements. The linear polarization resistance, $R_P$, was measured at an interval of $\pm 5$ mV concerning $E_{\rm corr}$. The Stern-Geary coefficient, B, was selected as $26$ mV/decade \cite{garcia2021experimental}.  

\section{Results and Discussion}
\subsection{Cyclic Voltammetry}
\begin{figure}[ht]
    \centering
    \includegraphics[width=\linewidth]{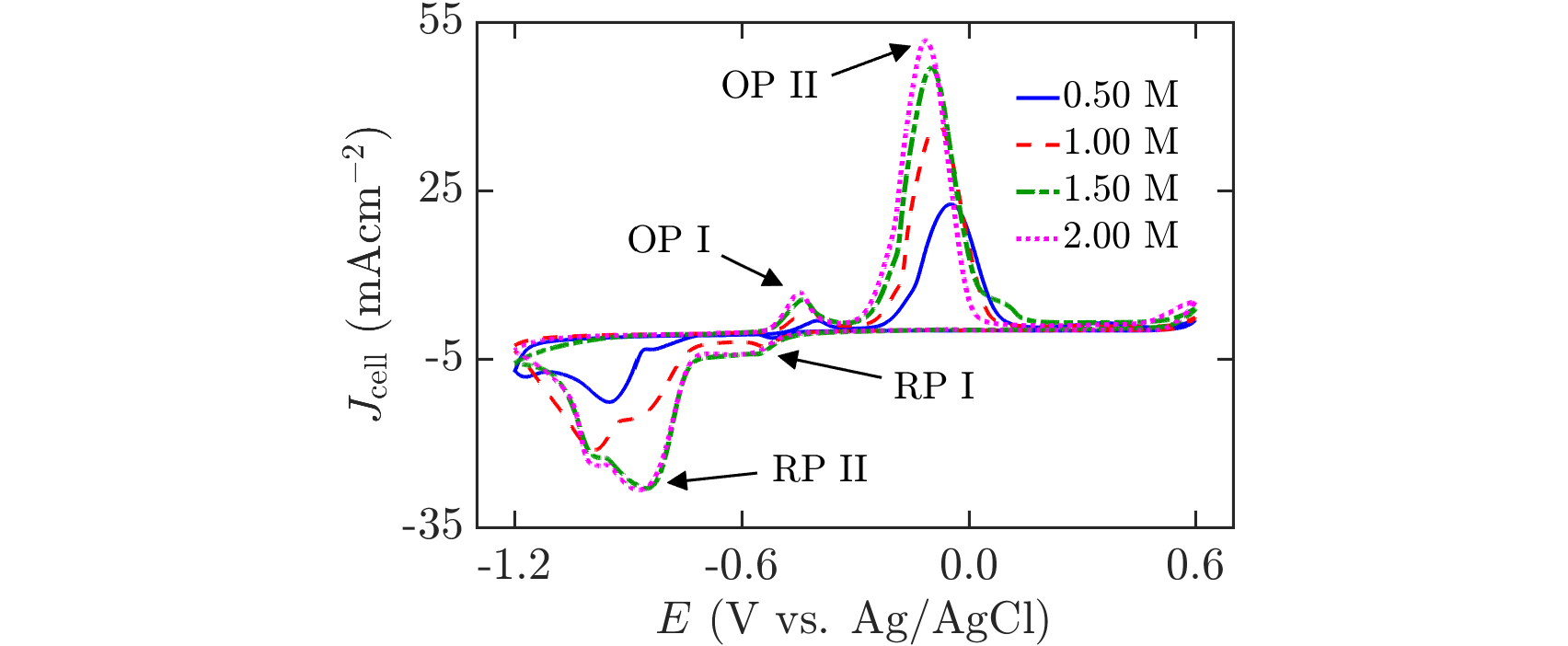}
    \caption{Cyclic Voltammetry (CV) of pure Cu in varying concentrations of KOH solution from 0.50 to 2.00 M. The oxidation peaks (OP) I and II represent the existence of Cu$^+$ and Cu$^{2+}$ ions, respectively. The reduction peaks (RP) I and II are attributed to the reduction of these two ions, respectively.}
    \label{Figure 03}
\end{figure}

The CV of Cu was measured at various concentrations of KOH solution, as shown in Fig.~\ref{Figure 03}. Four distinct peaks were observed in the CV curves. The two positive peaks were attributed to oxidation reactions, while the two negative peaks were present due to reduction reactions. The oxidation peak (OP) I indicated the formation of a Cu$_2$O monolayer, resulting from the oxidation of Cu through the following reaction \cite{eid2024surface}
\begin{equation}
    \rm{2Cu + 2OH^{-} \rightarrow Cu_2O + H_2O + 2e^{-}}.
    \label{eq:16}
\end{equation}
The peak shifted toward the left with increasing $C$, showing its dependence on the ionic strength of the electrolyte. At $C=0.5$ M, the peak was centered at $E=-0.40$ V vs.~Ag/AgCl, which agreed with the previously reported value \cite{giri2016electrochemical}. At $C=0.25$ M, the current of the OP I, $J_{\rm ox,I}$, was negligible. However, $J_{\rm ox,I}$ increased $3.67$ times when $C$ was increased to $2.00$ M from $0.50$ M. Despite the improvement, the tiny area of the peak represented a small amount of Cu$_2$O formation. Contrary to the OP I, the OP II was sharper and more prominent. The peak was located at $E=-0.05$ V vs.~Ag/AgCl, which agreed well with the value reported in the literature too \cite{giri2016electrochemical}. The oxidation peak II shifted toward lower voltages with increasing $C$. Two distinct reaction pathways contribute to the formation of this peak \cite{eid2024surface}, one of which is the direct two-electron oxidation of Cu, while the other reaction is the oxidation of the Cu$_2$O monolayer. The corresponding reactions can be written as \cite{eid2024surface}
\begin{subequations}
\begin{equation}
    \rm{2Cu + 2OH^{-} \rightarrow CuO + H_2O +2e^{-}},
    \label{eq:17(a)}
\end{equation}
\begin{equation}
    \rm{Cu_2O + 2OH^{-} \rightarrow 2CuO + H_2O +2e^{-}}.
    \label{eq:17(b)}
\end{equation}
\end{subequations}
The current observed at the OP II, $J_{\rm ox,II}$, was significantly higher than $J_{\rm ox,I}$, indicating that the formation of CuO is more favorable. The increase of $J_{\rm ox,II}$ was $>6$ times when $C$ rose from $0.25$ to $2.00$ M. 

The reduction peak (RP) I results from the reduction of CuO to Cu$_2$O. Similar to OP I and II, the reduction peaks shifted toward the left with increasing ionic strength of the electrolyte. The current density at RP I, $J_{\rm red,I}$, is smaller than all the other peaks present in the CV, showing that this reduction pathway is the least favorable. On the contrary, RP II is broader than RP I. This characteristic can be represented by any of the two reactions as follows: (i) the reduction of CuO to Cu; and (ii) the reduction of Cu$_2$O to Cu. The current density at RP II, $J_{\rm red,II}$, was significantly higher than $J_{\rm red,I}$ across all $C$. Table S1 shows the current densities of all four peaks for $C = 0.25$ to $2.00$ M.
\begin{figure}
    \centering
    \includegraphics[width=\linewidth]{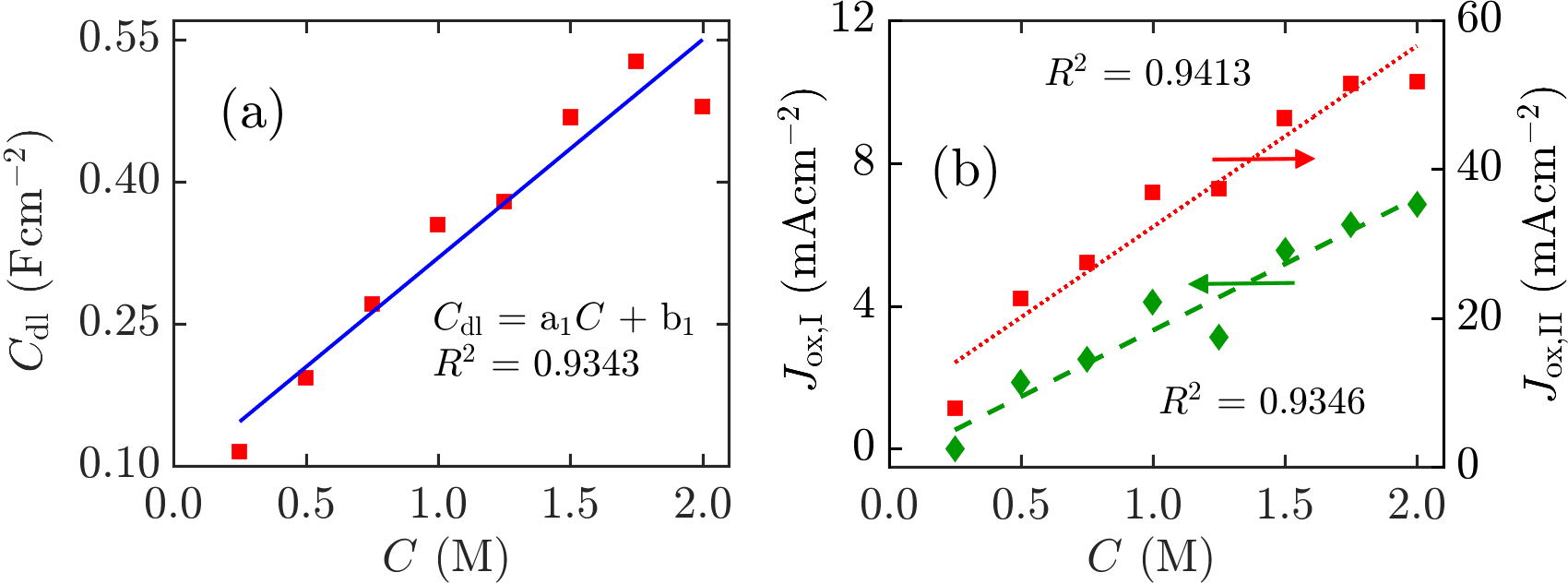}
    \caption{(a) Experimental data and regression line for the double layer capacitance ($C_{\rm dl}$) formed at the interface between Cu electrode and KOH solution at different concentrations ($C$). (b) Experimental data and regression lines for the oxidation peak current densities ($J_{\rm ox,I}$ and $J_{\rm ox,II}$) for the Cu electrodes in varying concentrations of KOH solution.}
    \label{Figure 04}
\end{figure}

Figure \ref{Figure 04}(a) shows a relationship between $C_{\rm dl}$ and $C$, and Table S1 represents the values of $C_{\rm dl}$ at different $C$. The higher $C$ enhances the ionic strength of the electrolyte, which generates stronger electric fields near the metal--electrolyte interface. Due to the screening effect of the electric field, the diffusion length of the ions, subsequently the Debye length, tends to get shorter. This results in the formation of thinner double layers and higher $C_{\rm dl}$. Lower double-layer thickness ensures less resistance for charge transfer. Due to the lower charge transfer resistance of the electrode--electrolyte interface at higher $C$, charge carrier transport becomes highly favorable, which leads to enhanced reaction kinetics for both HER and metal dissolution. In Fig.~\ref{Figure 04}(a), we also show a regression line. The regression line parameters, $a_1$ and $b_1$ had values of $0.2304$ Fcm$^{-2}$M$^{-1}$ and $0.0889$ Fcm$^{-2}$, respectively. The value of the coefficient of determination, $R^2$, was $0.9343$, demonstrating a strong correlation between $C_{\rm dl}$ and $C$. 

Figure \ref{Figure 03}(b) illustrates the relation of $J_{\rm ox,I}$ and $J_{\rm ox,II}$ with varying $C$. The regression equations for $J_{\rm ox,I}$ and $J_{\rm ox,II}$ were calculated as given by 
\begin{subequations}
    \begin{equation}
        J_{\rm ox,I} = a_2 C + b_2,
        \label{eq:18(a)}
    \end{equation}
    \begin{equation}
        J_{\rm ox,II} = a_3 C + b_3.
        \label{eq:18(b)}
    \end{equation}
\end{subequations}
The determination coefficient $R^2$ for the regression equations of $J_{\rm ox,I}$ and $J_{\rm ox,II}$ were at $0.9346$ and $0.9413$, respectively, demonstrating excellent agreement with experimental data. The regression line for $J_{\rm ox,1}$ was fitted using $a_2$ and $b_2$, having values of $3.717$ mAcm$^{-2}$M$^{-1}$ and $-0.4011$ mAcm$^{-2}$, respectively. The negative value of $b_2$ suggests that the existence of OP I is not possible at low C ($<0.1079$ M). However, the positive value of $b_3$ shows that the formation of the oxidation Peak II is quite feasible even at low $C$. The values of $a_3$ and $b_3$ parameters were calculated to be $24.34$ mAcm$^{-2}$M$^{-1}$ and $7.892$ mAcm$^{-2}$, respectively. 

\subsection{Corrosion Analysis}

\begin{figure}
    \centering
    \includegraphics[width=\linewidth]{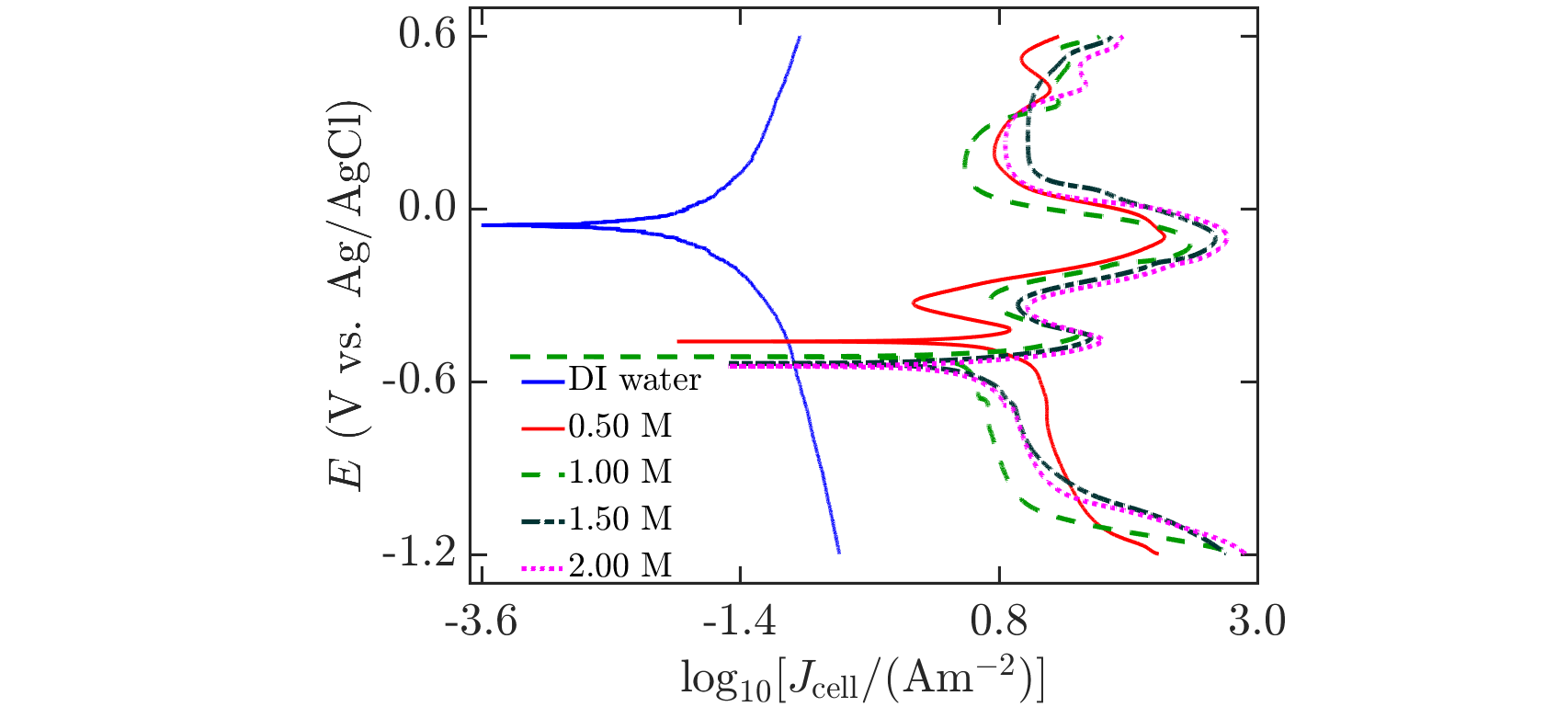}
    \caption{Potentiodynamic polarization measurement of pure Cu at DI water and varying concentrations of KOH. The concentration of KOH has been varied from $0.50$ to $2.00$ M. The room temperature was $25$ $^{\rm o}$C.}
    \label{Figure 05}
\end{figure}

The potentiodynamic polarization curves of Cu in DI water and various concentrations of alkaline solution are illustrated in Fig.~\ref{Figure 05}. The curves showed two regions, namely the anodic and cathodic regions. When $E$ was smaller compared to $E_{\rm corr}$, the cathodic region was observed. However, at $E>E_{\rm corr}$, the anodic region was observed, which contained two peaks. The peaks were similar to the oxidation peaks observed in the CV analysis. The peaks indicated the formation of metal oxides such as Cu$_2$O and CuO. Table S2 includes the values of $E_{\rm corr}$, $J_{\rm corr}$, the polarization resistance ($R_P$), and $C_R$ in DI water and $0.25$ to $2.00$ M KOH solution.

At DI water, the $E_{\rm corr}$ was recorded as $-0.05691$ V.~ This slightly negative $E_{\rm corr}$ indicated that the metal surface had a low tendency to lose electrons, showing high corrosion resistance. $R_P$ provides a quantitative measurement of corrosion resistance. The $R_P$ observed at DI water was more than $21$ k$\Omega$cm$^{-2}$, exhibiting high tolerance to electrochemical corrosion. The low values of $J_{\rm corr}$ and $C_R$ indicated the high stability of Cu in DI water. $E_{\rm corr}$ shifted to more negative values with increasing $C$ of KOH solution. $R_P$ also showed a decreasing trend with increasing $C$. Therefore, electrons can transfer to the electrolyte from the metal surface more easily due to the deterioration of the corrosion resistance. This behavior is further supported by the increasing trends of both $J_{\rm corr}$ and $C_R$. 

When $C$ increased from $0.25$ to $1.00$ M, both $J_{\rm corr}$ and $C_R$ increased by $1.9$ times. However, $E_{\rm corr}$ decreased only by a factor of $1.1$. A small change in $E_{\rm corr}$ caused significant changes in both $J_{\rm corr}$ and $C_R$. Because of the stronger electric field developed at the electrode--electrolyte interface at higher $C$, a large number of electrons present on the metal surface get energized. Moreover, the electric field enhances the drift velocity of electrons, leading to rapid electron transport. The kinetics of metal electrodissolution become more favorable due to the fast transport of electrons. As a result, the electrochemical corrosion of metals becomes more pronounced in high $C$. 

\subsection{Electrochemical Corrosion Model Fitting}

The experimental corrosion data of Cu electrodes (refer to Table S2) were used to fit the developed electrochemical corrosion model and calculate the empirical parameters of the governing Eqs.~\eqref{eq:12}--\eqref{eq:14}. $P$ was considered $1$ atm following our experimental conditions. Our regression analysis, discussed in section 4.1, suggested that the formation of the unstable oxidation peak was not possible below $\sim0.1$ M. Therefore, the model was fitted in two different conditions as follows: (i) $0\leq C<0.1$ M and (ii) $C\geq0.1$ M. Eq.~\eqref{eq:12} was considered as an objective function, which was optimized in MATLAB using the interior-point algorithm with a tolerance of $10^{-9}$. The empirical parameters, such as $\beta_{{\rm M},k}$, $\beta_{\rm H}$, $b_{{\rm M},k}$, and b$_{\rm H}$, were optimized by minimizing the squared error (SE) of the objective function. The lower and upper bounds for $\beta_{{\rm M},k}$ and $\beta_{\rm H}$ were selected as $1.1$ and $5$, respectively. Meanwhile, the bounds for $b_{\rm M,1}$, $b_{\rm M,2}$, and $-b_{\rm H}$ ranged between $0$ and $0.5$ V. 

\begin{figure}
    \centering
    \includegraphics[width=\linewidth]{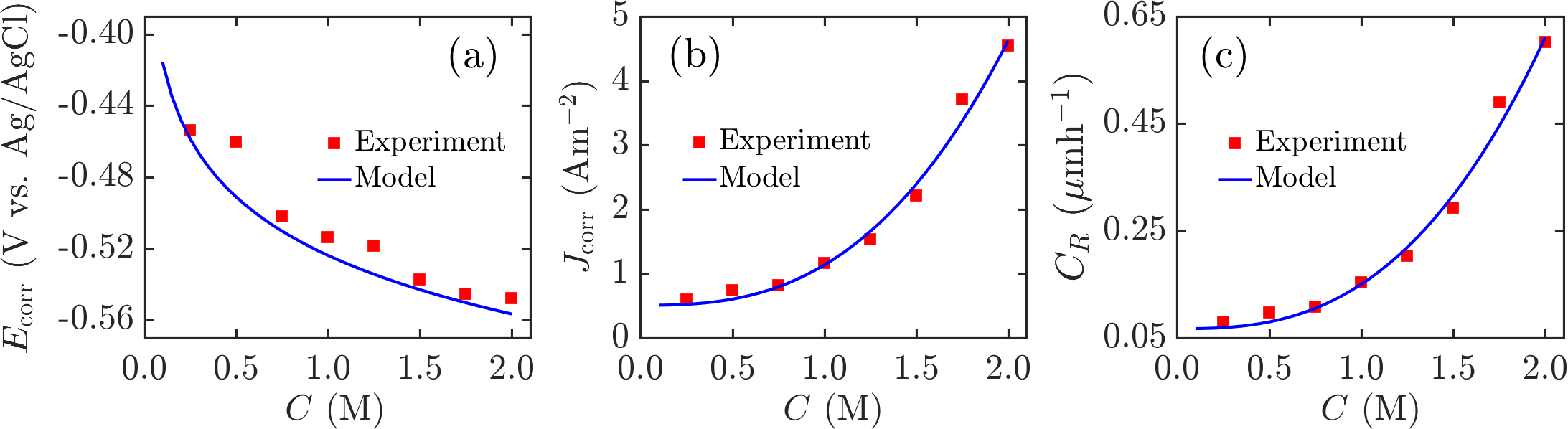}
    \caption{Comparison between the experimental and theoretical values of (a) corrosion potential ($E_{\rm corr}$), (b) corrosion current density ($J_{\rm corr}$), and (c) corrosion rate ($C_R$).}
    \label{Figure 06}
\end{figure}

At $C\geq0.1$ M, Cu exhibits only two oxidation states in KOH, allowing us to consider $L$ as $2$. $k=1$ corresponded to the stable oxidation state of $+2$, whereas $k=2$ indicated the unstable $+1$ oxidation state. The optimization process was carried out for $C$ ranging from $0.25$ to $2.00$ M. The ranges of values of $\beta_{\rm M,1}$, $\beta_{\rm M,2}$, and $\beta_{\rm H}$ were obtained as $2.39$--$2.80$, $2.29$--$2.95$, and $1.79$--$2.81$, respectively. Meanwhile, $b_{\rm M,1}$, $b_{\rm M,2}$, and $-b_{\rm H}$ were calculated as $0.2497$--$0.2528$, $0.2577$--$0.2628$, and $0.0469$--$0.0604$ V, respectively. The parameters were further optimized to achieve the best possible fit with the experimental data. Using a similar approach, the values of $\beta_{\rm M,1}$, $\beta_{\rm H}$, b$_{\rm M,1}$, and b$_{\rm H}$ were optimized for the range $0\leq C<0.1$. The fitted parameters for both cases are summarized in Table \ref{Table:01}. Using the fitted parameters of Table \ref{Table:01}, Eqs.~\eqref{eq:12} and \eqref{eq:13} were numerically solved in MATLAB to find $E_{\rm corr}$, $J_{\rm corr}$, and $C_R$ for $C$ ranging from $0.1$ to $2.0$ M. The theoretical and experimental values of these corrosion parameters are illustrated in Fig.~\ref{Figure 06}. 

\begin{longtable}{p{1.02in} p{0.5in} p{0.5in} p{0.5in} p{0.59in} p{0.59in} p{0.59in}}
\caption{Summary of fitted parameters for the advanced electrochemical corrosion model. \label{Table:01}} \\
\hline \hline
Region & $\beta_{\rm M,1}$ & $\beta_{\rm M,2}$ & $\beta_{\rm H}$ & $b_{\rm M,1}$ (V) & $b_{\rm M,2}$ (V) & $b_{\rm H}$ (V) \\
\hline
$0\leq C<0.1$ M & $1.985$ & $-$ & $1.52$ & $0.01928$ & $-$ & $-0.0069$ \\
$C\geq0.1$ M & $2.7951$ & $2.9478$ & $2.3121$ & $0.2527$ & $0.2628$ & $-0.0537$ \\
\hline \hline
\end{longtable}

Statistical analyses, such as the mean absolute percentage error (MAPE) and the standard deviation ($\sigma$), were calculated to assess the correlation between the model and the experimental data. These results are summarized in Table \ref{Table:02}. Figure \ref{Figure 06}(a) suggested that the model consistently underpredicted $E_{\rm corr}$. However, the MAPE remained within $\sim 5\%$, showing excellent correlation between the model and experimental data. Meanwhile, the model did not exhibit systematic trends of under-prediction or over-prediction in Figs.~\ref{Figure 06}(b) and \ref{Figure 06}(c). The MAPE for both $J_{\rm corr}$ and $C_R$ remained within $10\%$, indicating a strong correlation between theory and experiment. Furthermore, the small $\sigma$ for all the parameters demonstrated the accuracy and reliability of the model. 
\begin{longtable}{p{1.2in} p{0.6in} p{0.6in} p{0.6in}}
\caption{Statistical evaluation of mean absolute percentage error (MAPE) and standard deviation ($\sigma$) for the developed advanced electrochemical corrosion model. \label{Table:02}} \\
\hline \hline
Statistical metric & $E_{\rm corr}$ & $J_{\rm corr}$ & $C_R$ \\
\hline
MAPE & $3.6194\%$ & $8.9761\%$ & $8.9761\%$ \\
$\sigma$ & $0.0083$ & $0.1032$ & $0.0137$ \\
\hline \hline
\end{longtable}

Some cautions should be maintained when using the model for assessing the electrochemical corrosion of metals other than Cu. $\beta_{\rm M,1}$ should be selected for the current component due to the most stable oxide. However, $\beta_{\rm M,2}$ must be chosen for the relatively unstable oxides. $b_{\rm M,1}$ and $b_{\rm M,2}$, shown in Table \ref{Table:01}, were calculated for two and one electron transfers, respectively. Eq.~\eqref{eq:3(a)} provided a relationship where $b_a$ is inversely proportional to $z$. This relationship can be utilized to calculate $b_{\rm M,1}$ and $b_{\rm M,2}$ for different numbers of electron transfers. Unlike Cu, some metals, such as Ag and Al, lose one and three electrons, respectively, to obtain their stable oxidation states. The value of $b_{\rm M,1}$ will be $0.5054$ V for Ag and $0.1685$ V for Al at $C\geq 0.1$ M. Similar analysis at $0\leq C<0.1$ M provides the value of $b_{\rm M,1}$ as $0.0386$ and $0.0129$ V for one and three electron transfers, respectively. To obtain unstable oxidation states, metals, such as Ag and Ni, give up two and three electrons, respectively. Therefore, the value of $b_{\rm M,2}$ will become $0.1314$ V for Ag and $0.0876$ V for Ni.

\subsection{Applications of the Advanced Electrochemical Corrosion Model}

\begin{figure}[ht]
    \centering
    \includegraphics[width=\linewidth]{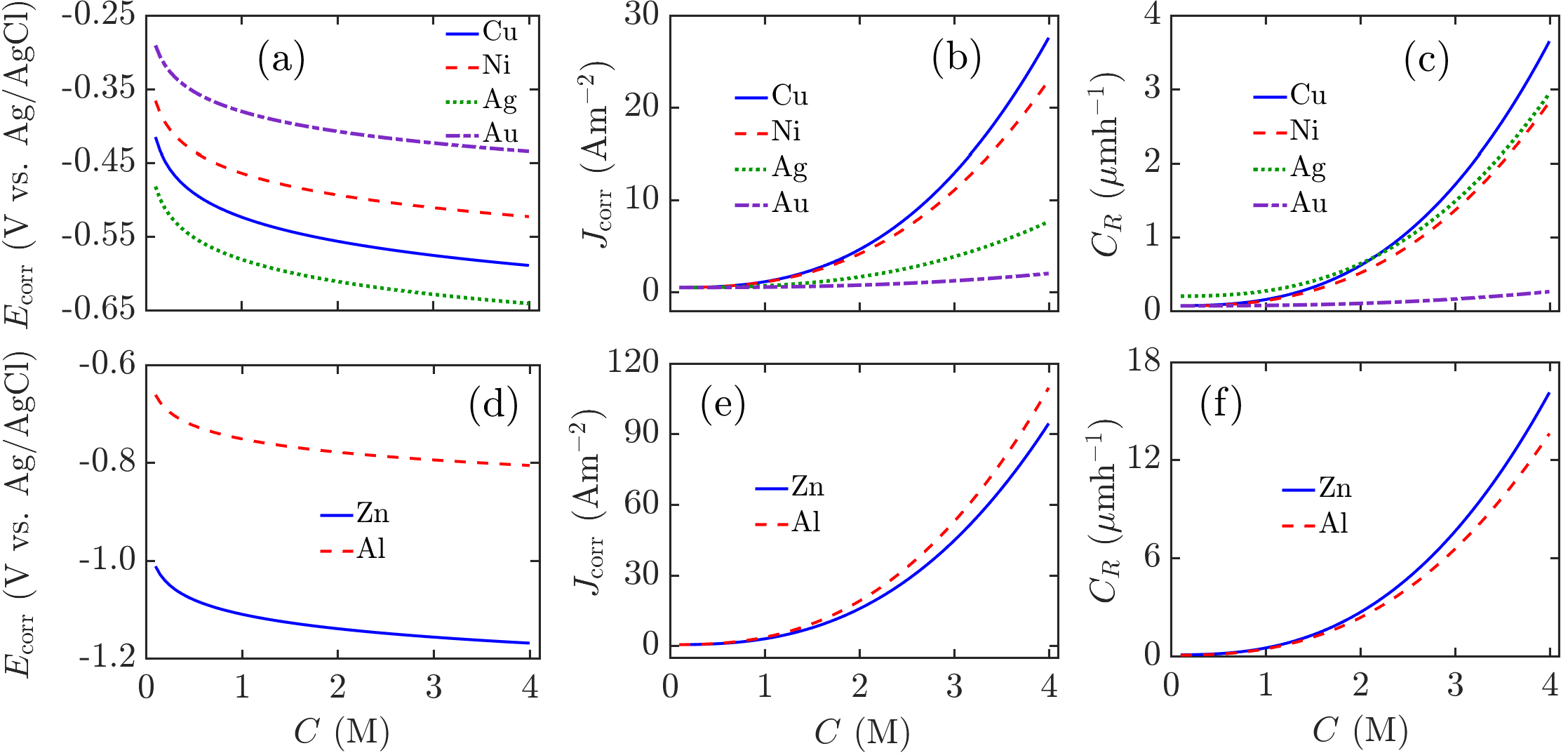}
    \caption{(a) The corrosion potential ($E_{\rm corr}$), (b) corrosion current density ($J_{\rm corr}$), and (c) corrosion rate ($C_R$) of Cu, Ni, Ag, and Au in varying concentrations of KOH. (d) The $E_{\rm corr}$, (e) $J_{\rm corr}$, and (f) $C_R$ of Zn and Al in varying concentrations of KOH.}
    \label{Figure 07}
\end{figure}

The developed model was used to measure the $E_{\rm corr}$, $J_{\rm corr}$, and $C_R$ of conventional metal electrodes used in alkaline water electrolyzers, including Cu, Ni, Ag, Au, Zn, and Al. The corrosion behaviors of these metals in monohydroxy electrolyte of varying $C$ from $0.1$ to $4.0$ M are illustrated in Fig.~\ref{Figure 07}. The value of $L$, in Eqs.~\eqref{eq:12} and \eqref{eq:13}, was selected for each metal, depending on the number of oxidation states. Ni exhibits two oxidation states in an alkaline electrolyte, such as $+2$ and $+3$ in the form of NiO and NiOOH, respectively \cite{hall2014oxalate}. Ag has a stable oxidation state of $+1$ and an unstable oxidation state of $+2$ in alkaline electrolytes \cite{luo2020investigation}. Therefore, the value of $L$ was two for these metals. Since Au, Zn, and Al exhibited a single oxidation state in alkaline electrolytes, the value of $L$ was considered as one \cite{cherevko2014comparative, trudgeon2019screening, abdallah2016gelatin}

The $E_{\rm corr}$ of all the metals exhibited a decreasing trend, while both $J_{\rm corr}$ and $C_R$ showed an exponentially increasing trend with increasing $C$. These trends are illustrated in Fig.~\ref{Figure 07}. The $E_{\rm corr}$ of Ni was calculated as $-0.54$ V vs. Ag/AgCl in $6.0$ M KOH. Meanwhile, the $E_{\rm corr}$ of Cu in $1.0$ M KOH solution was calculated as $-0.5237$ V vs.~Ag/AgCl. The model predictions for both Cu and Ni showed excellent agreement with experimental data \cite{syam2023combination,arnaudova2024corrosion}. The $E_{\rm corr}$ of Zn, according to the model, was $-1.157$ V vs.~Ag/AgCl at $3.0$ M KOH, which agreed well with previously reported data \cite{li2012composite}. 

Noble metals, such as Au and Ag, showed smaller $J_{\rm corr}$ than both Cu and Ni. However, the $C_R$ of Ag exceeded that of Cu up to $\sim 2.1$ M. According to Eq.~\eqref{eq:15}, $C_R$ depends on the ratio between $M$ and $z_{\rm M}d$, which includes the intensive properties of a metal. This ratio represents the volume of metal lost per mole of electrons transferred during the corrosion process. For Ag, this ratio was more than $3$ times greater than the other metals. Additionally, the work function of Ag is less than both Cu and Ni, which makes the removal of electrons more favorable \cite{michaelson1977work}. Therefore, Ag showed a higher $C_R$ than both Cu and Ni at low $C$. However, due to the exponential nature of $J_{\rm corr}$, the $C_R$ of Cu exceeded that of Ag after $C\sim2.1$ M. Furthermore, the $C_R$ of Ag and Ni were nearly equal in this range. According to the electrochemical corrosion model, Au showed negligible $C_R$, making it the most stable metal for alkaline electrolysis. The high work function of Au makes the removal of the 6s electron energetically unfavorable \cite{abas2014electrode}. Additionally, Au does not oxidize to form metal-hydroxide complexes in alkaline electrolytes due to its weak chemical bond with OH$^-$ ions. Therefore, Au shows a high degree of electrochemical inertness in an alkaline environment. 

Although the work functions of both Zn and Al are nearly equal, their corrosion characteristics are different \cite{michaelson1977work}. Their corrosion parameters are illustrated in  Figs.~\ref{Figure 07}(d)--\ref{Figure 07}(f). Despite showing a higher $J_{\rm corr}$ than Zn, Al had a lower $C_R$. The $J_{\rm corr}$ of Al was at most $1.2253$ times higher than that of Zn. However, the ratio between $M$ and $z_{\rm M}d$ for Zn was $1.3745$ times larger than that of Al. As a result, Zn corroded faster despite having a lower $J_{\rm corr}$. 

\begin{figure}
    \centering
    \includegraphics[width=\linewidth]{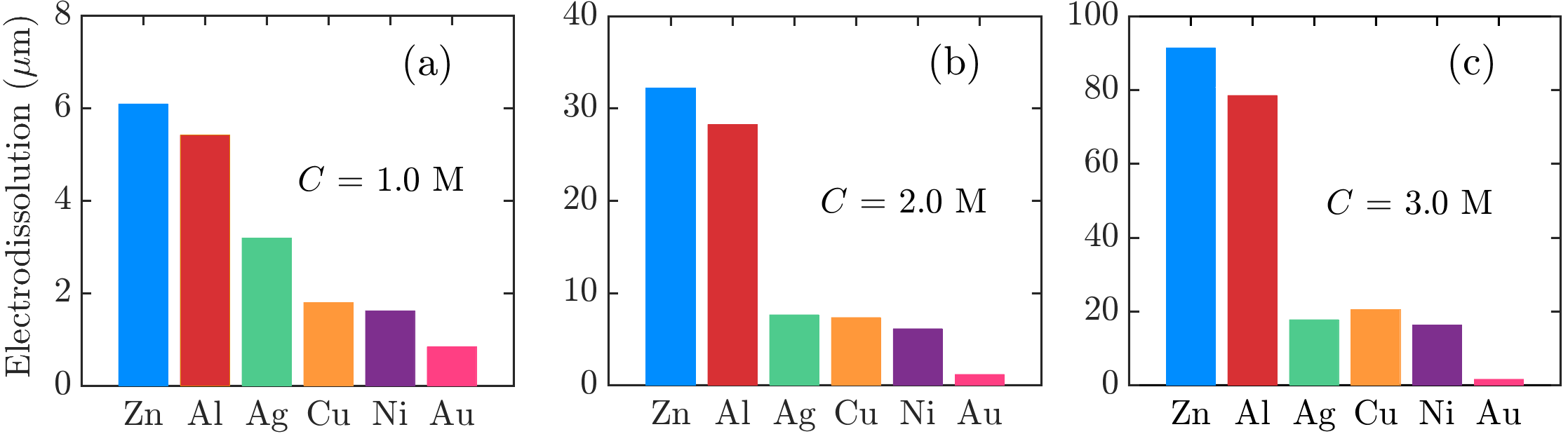}
    \caption{Electrodissolution of Zn, Al, Cu, Ag, Ni, and Au after 12 hours of exposure to (a) $1.0$, (b) $2.0$, and (c) $3.0$ M KOH solution.}
    \label{Figure 08}
\end{figure}

The model can be used to assess the long-term stability of the metal electrodes by measuring their electrodissolution. Figure \ref{Figure 08} shows the electrodissolution of Zn, Al, Ag, Cu, Ni, and Au after $12$ hours of operation in alkaline solution at $C$ of $1.0$, $2.0$, and $3.0$ M. Since Zn and Al dissolved at a higher rate than all the other metals in all $C$, they are the most unstable metal electrodes to use in alkaline water electrolyzers. On the contrary, Cu and Ni dissolved $4.47$ and $5.60$ times less than Zn in $3.0$ M KOH solution, respectively. The low electrochemical dissolutions of Cu and Ni make them promising electrode materials for alkaline water splitting. However, noble metal electrodes are highly desirable for efficient and sustainable H$_2$ generation. Despite being a noble metal, Ag was less stable than Au. In low electrolytic strength, Ag exhibited greater dissolution than Cu and Ni. However, Ag dissolved less than Cu and nearly as much as Ni at $C=3.0$ M. Therefore, Ag can be utilized as a stable electrode only in alkaline electrolytes of high ionic strength. The electrodissolution of Au was $\sim48$ times smaller than that of Zn in $3.0$ M KOH solution. Au showed outstanding stability because of its extremely low dissolution, indicating that Au was the most desirable candidate for designing efficient and sustainable alkali-based water electrolyzers.

\section{Conclusion}
Water electrolysis is a leading technology for producing green H$_2$ fuel. Alkaline water electrolyzers have become particularly attractive for renewable H$_2$ generation due to their commercial availability. However, electrochemical corrosion poses a significant challenge that reduces the overall efficiency and durability of these devices. In this study, we developed an advanced physics-informed electrochemical corrosion model to investigate the corrosion behavior of metal cathodes used in standard water electrolyzers. The model incorporates MPT, material properties, and system parameters such as concentration ($C$), pressure ($P$), and temperature ($T$). This allowed us to analyze key corrosion metrics, including $E_{\rm corr}$, $J_{\rm corr}$, $C_R$, and the electrodissolution of metal electrodes in various electrochemical environments. The reliability and validity of the model were confirmed through statistical evaluations. Our analysis of the long-term stability of several conventional metal electrodes revealed that Au, copper Cu, and Ni are the most durable options for alkaline water electrolyzers. The insights gained from this model are valuable for designing more resilient alkaline water electrolyzers.
%
\section*{Supplementary Material}
The supplementary material contains the impact of $\gamma_{{\rm M}}$ and $\gamma_{\rm H}$ variation on $J_{\rm corr}$, and the experimental data obtained from electrochemical analyses such as cyclic voltammetry and potentiodynamic polarization. 

\section*{Data Availability}
All data of the paper are presented in the main text and the supplementary material.

\section*{Author Declaration}
The authors have no conflicts of interest to disclose.
  
\small
\bibliographystyle{ieeetr}
\bibliography{references}
\end{document}